# Analytical study of superconducting RF cavity detuning compensation

Faya Wang and Chris Adolphsen

**Abstract** The cavities in superconducting linacs tend to operate with narrow bandwidths that are comparable to microphonic induced detuning from external forces. Proportional-Integral (PI) feedback loops are used to stabilize the cavity voltage in response to the varying cavity resonant frequency. Without feedback, the cavity voltage variation with detuning is non-linear, but in this paper we show that in the regime where the detuning amplitude is smaller than the frequency of the detuning, which is typically the case, the response can be linearized so as to provide simple analytical characterizations of PI loop performance as a function of the various parameters involved (integral and proportional gains, noise and low pass filter bandwidth).

**Introduction**

Superconducting linacs for XFELs and ERLs that operate CW can in principle have very narrow cavity bandwidths (BWs) for low current operation, which reduces the required RF power to the minimum needed to accelerate the beam. In practice, the cavity BW chosen in this case is often constrained by the microphonics level, that is, vibration and He pressure induced detuning of the cavities, which requires RF overhead to compensate. For the LCLS-II Linac, for example, the cavity half BW was chosen to be comparable to the maximum expected detuning. This makes stabilizing the cavity gradients challenging, especially as tight gradient regulation is required ($10^{-4}$ level). As is typical, gradient regulation in LCLS-II will be achieved using Proportional-Integral (PI) feedback loops, which stabilize the signals measured from probes of the cavity fields. In this paper, we present a linearized analytical analysis of the feedback control in the presence of microphonics and derive simple expressions for the gradient and phase stability achievable based on the loop gain parameters, electronic noise and the detuning spectrum. We also calculate the associated RF power overhead required.

**Cavity Equations**

The parallel LC circuit shown in Fig. 1 was used to model an RF and beam driven cavity whose resonant frequency varies with time. The cavity voltage (i.e., the voltage across this circuit) can be derived from the following time dependent circuit equation:

$$C \frac{d^2 \tilde{V}}{dt^2} + \left(\frac{1}{R_L} + 2\frac{dC}{dt}\right)\frac{d\tilde{V}}{dt} + \left(\frac{d^2 C}{dt^2} + \frac{1}{L}\right)\tilde{V} = \frac{d\tilde{I}_s}{dt}, \qquad (1)$$

where $\tilde{I}_s = \tilde{I}_g + \tilde{I}_b$ is the sum of generator (RF source) and beam currents, $\tilde{V}$ is the cavity voltage, $R_L$ is the loaded shunt impedance, i.e., $\frac{1}{R_L} = \frac{1}{R_g} + \frac{1}{R}$. For this analysis, we assume the beam current is zero ($\tilde{I}_b = 0$), and the only variation in the drive power is to stabilized the cavity voltage in the presence of detuning. Given that the voltage can be made very stable (typically $< 10^{-3}$), we ignore the effect of Lorentz force detuning.

If $\omega_0$ is the nominal angular frequency ($\omega_0 = 2\pi f_0$) and $Q$ is unloaded quality factor, we assume $L$ and $C$ vary with time as

$$L = \frac{R/Q}{\omega_0}[1 - \varepsilon_l(t)], \tag{2a}$$

$$C = \frac{1}{(R/Q)\omega_0}[1 - \varepsilon_c(t)]. \tag{2b}$$

where $\varepsilon_l$ and $\varepsilon_c$ are $\ll 1$, that is, the time variation of the detuning is many orders of magnitude smaller than the cavity frequency. In this case, $\omega = \omega_0[1 + \varepsilon(t)]$ is a good approximation where $\varepsilon = (\varepsilon_l + \varepsilon_c)/2$.

Expressing the voltage as $\tilde{V} = Ve^{j\omega_0 t}$ and current as $\tilde{I}_s = I_s e^{j\omega_0 t}$, the derivatives of $\tilde{V}$ and $\tilde{I}_s$ are:

$$\frac{d\tilde{V}}{dt} = \left(j\omega_0 V + \frac{dV}{dt}\right)e^{j\omega_0 t}, \tag{3a}$$

$$\frac{d^2\tilde{V}}{dt^2} = \left(-\omega_0^2 V + j2\omega_0 \frac{dV}{dt} + \frac{d^2V}{dt^2}\right)e^{j\omega_0 t}, \tag{3b}$$

$$\frac{dI_s}{dt} = \left(j\omega_0 I_s + \frac{dI_s}{dt}\right)e^{j\omega_0 t}, \tag{3c}$$

Substituting Eq. (2) and (3) into (1) assuming $Q_L = Q\,R_L/R$ is large (typically $> 10^7$) and neglecting the terms that are second order or higher in the slow (typically $< 1$ kHz) variation of the fields and detuning, yields

$$\dot{V} + (1 - j\delta)V = V_g \tag{4}$$

where the derivative is with respect to $\tau = \frac{t}{T_F}$ with $T_F = \frac{2Q_L}{\omega_0}$, the time constant to fill the cavity. Also, $V_g = I_s R_L$ and $\delta = \frac{d}{B}$, that is, the cavity detuning, $d = \varepsilon f_0$, divided by $B = \frac{f_0}{2Q_L}$, the cavity half bandwidth. Thus to fully compensate the field fluctuation (i.e., $\dot{V} = 0$), the generator voltage needs to be $(1 - j\delta)V$ and the fraction of additional power needed is about $\delta^2$, which is independent of the frequency of microphonics.

**Linearization**

With the cavity initially unfilled, the cavity voltage evolves when driven by a step change in $V_g$ as

$$V(\tau) = e^{-\int_0^\tau (1-j\delta)d\eta} \int_0^\tau e^{\int_0^\zeta (1-j\delta)d\eta} V_g d\zeta, \tag{5}$$

which is derived from integrating Eq. (4). We now consider the case where the detuning occurs at a single frequency,

$$\delta = \delta_0 e^{j\beta_0 \tau}, \tag{6}$$

where $\delta_0$ is the detuning amplitude (complex number) normalized by $B$ and $\beta_0$ is the frequency of the detuning, also normalized by $B$. Eq. (5) then simplifies to

$$V(\tau) = e^{-\tau - \frac{\delta_0}{\beta_0}(1-e^{j\beta_0\tau})} \int_0^\tau e^{\zeta - \frac{\delta_0}{\beta_0}(1-e^{j\beta_0\zeta})} V_g d\zeta. \qquad (7)$$

This equation shows that the cavity voltage depends nonlinearly on the detuning, which produces voltage variations at the harmonics of the detuning frequency for constant $V_g$. However, for $\delta_0/\beta_0 \ll 1$, that is, when the detuning amplitude is much smaller than the detuning frequency,

$$V(\tau) \approx e^{-\tau}\left[1 - \frac{\delta_0}{\beta_0}(1 - e^{j\beta_0\tau})\right] \int_0^\tau e^\zeta \left[1 + \frac{\delta_0}{\beta_0}(1 - e^{j\beta_0\zeta})\right] V_g d\zeta, \qquad (8)$$

which varies at only the detuning frequency to first order in $\delta_0/\beta_0$ for constant $V_g$. To compensate the detuning in this case, we assume the generator voltage is modulated as,

$$V_g = V_{g0}(1 + \gamma e^{j\beta_0\tau}). \qquad (9)$$

The cavity voltage is then

$$\frac{V}{V_{g0}} \approx 1 + \left(\frac{\gamma - \delta_0/\beta_0}{1+j\beta_0} + \frac{\delta_0}{\beta_0}\right)e^{j\beta_0\tau} - \left(1 + \frac{\gamma - \delta_0/\beta_0}{1+j\beta_0} + \frac{\delta_0}{\beta_0}e^{j\beta_0\tau}\right)e^{-\tau}, \qquad (10)$$

and the second term is zero when $\gamma = -j\delta_0$. Thus, modulating the applied voltage out-of-phase by the detuning amplitude produces a constant cavity voltage in steady state (i.e., large $\tau$) to first order in $\delta_0/\beta_0$. This is in fact an exact steady state solution of Eq. 4. For $\gamma = 0$, the steady state voltage variation is

$$\frac{V}{V_{g0}} \approx 1 + \frac{j\delta_0}{1+j\beta_0}e^{j\beta_0\tau}, \qquad (11)$$

so the cavity voltage variation decreases as $1/\beta_0$ for large $\beta_0$.

To illustrate the accuracy of this first order expansion, Fig. 2 shows a comparison of the cavity voltage computed by numerical integration of Eq. (7) with that computed from Eq. (10) for various $\delta_0$ and $\beta_0$ values with $\gamma = 0$ (i.e., the generator turn-on is a step function). The deviation of the linear approximation from the rigorous solution decreases with smaller $\delta_0/\beta_0$ and with smaller $\delta_0$ as illustrated in Fig. 3. For $\delta_0/\beta_0 \lesssim 0.5$ and $\delta_0 \lesssim 0.5$, which is expected at LCLS-II, the approximation is better than about 10%.

**Feedback System**

A typical PI loop feedback system for a SRF cavity is illustrated in Fig. 4. The corresponding open and closed loop transfer functions (i.e., $V/V_s$) for the system are, respectively,

$$H(s) = \left(K_p + \frac{K_i}{s}\right)\frac{1}{1+s/2\pi L}\frac{1}{1+s/2\pi B}, \qquad (12)$$

$$T(s) = \frac{H(s)}{1+e^{-sD}H(s)}, \qquad (13)$$

where $s = i\omega$, $K_p$ and $K_i$ are the proportional and integral loop gain coefficients, $D$ is the round trip signal delay time and $L$ is the low pass filter bandwidth, which is assumed to be much larger than the half cavity bandwidth, $B$. Typically, $K_p$ is greater than $10^3$ and $K_i$ is comparable or larger than $2\pi B K_p$ to enhance the gain at low frequency (i.e., within the cavity bandwidth). Also $D$ is generally small enough ($<$ few $\mu s$) that $e^{-sD}$ can assumed to be unity for $H(s) > 1$.

An important consideration is the stability of the feedback loop, which is characterized the by phase margin (PM). This is the phase of $H$ relative to -180 degrees at the frequency where the open loop gain is unity. A large phase margin (PM) is preferred, which sets the scale of $K_i$. For $K_i = 2\pi B K_p$, the phase of $H(s)$ decreases smoothly from -90 degrees at low frequency to -180 degrees at high frequency for all $L$, and for this study, we assume this relation between $K_i$ and $K_p$. As a result, the zero crossing frequency (i.e., where $H = 0\ dB$) is

$$f_c \approx L \sqrt{\sqrt{\left(K_p \frac{B}{L}\right)^2 + \frac{1}{4}} - \frac{1}{2}}. \tag{14a}$$

This approximation agrees well with numeric results as shown in Fig. 5. For $K_p B / L \gg 1$,

$$f_c \approx \sqrt{K_p B L}. \tag{14b}$$

With $f_c \gg B$, $\varphi[H(2\pi f_c)] \approx -\pi + \tan^{-1}\frac{L}{f_c}$ so

$$PM = \tan^{-1}\frac{L}{f_c}. \tag{15}$$

**Feedback Compensation**

The full set of equations for the cavity plus feedback system and instrumentation noise without beam is,

$$\dot{v} + (1 - j\delta)v = v_g, \tag{16a}$$

$$\dot{v}_1 = K_p[\dot{n}_i - \dot{v}] + \frac{K_i}{2\pi B}[1 + n_i - v], \tag{16b}$$

$$\frac{B}{L}\dot{v}_g + v_g = v_1, \tag{16c}$$

where v, $v_1$ and $v_g$ are the voltages normalized to the set point voltage $V_s$. We express the detuning and instrument noise as sum of sinusoidal contributions:

$$\delta = \sum_{n=1}^{\infty} \delta_n e^{j\beta_n \tau}, \tag{17}$$

$$n_i = \sum_{n=1}^{\infty} n_{in} e^{j\beta_n \tau}, \tag{18}$$

where

$$n_{in} = \int_{\beta_n}^{\beta_{n+1}} 10^{N_i(\beta)/10} B d\beta, \quad (19)$$

and $N_i$ (in dBc/Hz) is the instrument noise power spectrum. To obtain an accurate linear solution to Eq. 16 in the time domain, we assume $\delta_n/\beta_n \lesssim 0.5$ and $\delta_n \lesssim 0.5$ as discussed above. The resulting steady state solution is

$$v = 1 + \sum_{n=1}^{\infty} a_n e^{j\beta_n \tau}, \quad (20)$$

$$v_g = 1 + \sum_{n=1}^{\infty} c_n e^{j\beta_n \tau}, \quad (21)$$

where

$$a_n = a_{dn} + a_{Nn}, \quad (22)$$

$$c_n = c_{dn} + c_{Nn}, \quad (23)$$

$$a_{dn} = -\frac{\beta_n(1+j\beta_n B/L)}{j\beta_n K_p + K_i/2\pi B - \beta_n^2(1+j\beta_n B/L)} \delta_n, \quad (24)$$

$$a_{Nn} = \frac{j\beta_n K_p + K_i/2\pi B}{j\beta_n K_p + K_i/2\pi B - \beta_n^2(1+j\beta_n B/L)} n_{in}, \quad (25)$$

$$c_{dn} = j\beta_n a_{dn} - j\delta_n, \quad (26)$$

$$c_{Nn} = j\beta_n a_{Nn}. \quad (27)$$

The subscript $dn$ and $Nn$ refer to the contribution from detuning and instrument noise, respectively. Since the noise and detuning spectral components are uncorrelated,

$$\sigma_v^2 \approx \sum_{n=1}^{\infty} \frac{|a_{dn}|^2}{2} + \sum_{n=1}^{\infty} \frac{|a_{Nn}|^2}{2}, \quad (28)$$

$$\sigma_{v_g}^2 \approx \sum_{n=1}^{\infty} \frac{|c_{dn}|^2}{2} + \sum_{n=1}^{\infty} \frac{|c_{Nn}|^2}{2}, \quad (29)$$

where $\sigma$ denotes the standard deviation. Thus RF overhead is also required as a result of the instrument noise, which is fed back as well and ultimately limits the cavity voltage stability. We will first discuss the detuning compensation.

Typically the dominate detuning is at frequencies below few hundred Hz while the low pass filter bandwidth is above a few kHz, so $\beta_n B/L \ll 1$. With $K_i = 2\pi B K_p$ and $\beta_n \ll K_p$,

$$a_{dn} \approx -\frac{\beta_n}{1+j\beta_n} \frac{\delta_n}{K_p}, \quad (30)$$

$$c_{dn} \approx -j\delta_n. \quad (31)$$

Consequently,

$$\sum_{n=1}^{\infty} \frac{|a_{dn}|^2}{2} \approx \frac{1}{K_p^2} \sum_n \frac{\beta_n^2}{1+\beta_n^2} \frac{\delta_n^2}{2}, \quad (32)$$

$$\sum_{n=1}^{\infty}\frac{|c_{dn}|^2}{2} \approx \sum_n \frac{\delta_n^2}{2}. \tag{33}$$

These approximations are in good agreement with numerical results as illustrated in Fig.6 where no instrument noise is included.

As another check, we used actual detuning data. Fig. 7a is a detuning spectrum measured for a LCLS-II cavity in a prototype cryomodule [1], and Fig. 7b shows that the simulated generator voltage corresponding to this spectrum using the approximations agrees well with the numeric solution. Given that $\beta_n \gtrsim 1$ for this spectrum, $a_{dn} \approx j\frac{\delta_n}{K_p}$ so the closed loop cavity voltage stability is

$$\sigma_v = \sqrt{\sum_{n=1}^{\infty}\frac{|a_{dn}|^2}{2}} \approx \frac{\sigma_d}{K_p}, \quad \text{where} \tag{34}$$

$$\sigma_d = \frac{\sqrt{\sum_n \delta_n^2}}{2}. \tag{35}$$

For the spectrum in Fig. 7a, $\sigma_d = 0.302$, and $K_p \sigma_V$ computed numerically is nearly independent of $K_p$ and equals $\sigma_d$ to within about 15 % as seen Fig. 7c.

Next we consider the noise contribution. As instrument noise is very broadband, $\beta_n$ is generally much larger than unity, for which

$$a_{Nn} \approx \frac{K_p}{K_p + j\beta_n(1+j\beta_n B/L)} n_{in}. \tag{36}$$

Therefore,

$$\sum_{n=1}^{\infty}\frac{|a_{Nn}|^2}{2} = \int_0^{\infty} 10^{N_i(\beta)/10} B \frac{K_p^2}{\left(K_p - \beta^2\frac{B}{L}\right)^2 + \beta^2} d\beta, \tag{37}$$

$$\sum_{n=1}^{\infty}\frac{|c_{Nn}|^2}{2} = \int_0^{\infty} 10^{N_i(\beta)/10} B \frac{K_p^2 \beta^2}{\left(K_p - \beta^2\frac{B}{L}\right)^2 + \beta^2} d\beta, \tag{38}$$

As expected, Eq. 37 is just the integral of the noise weighted by the square of the closed looped transfer function. Fig. 8 shows plots of the weighting functions in Eq. 37 and Eq. 38 for a phase margin of 20 degrees. For white noise (i.e., flat frequency spectrum), the above integrations yield

$$\sum_{n=1}^{\infty}\frac{|a_{Nn}|^2}{2} = \frac{\pi}{2} 10^{N_i/10} B K_p, \tag{39}$$

$$\sum_{n=1}^{\infty}\frac{|c_{Nn}|^2}{2} = \frac{\pi}{2} 10^{N_i/10} L K_p^2. \tag{40}$$

Combing the noise and detuning contributions yields,

$$\sigma_v^2 = \frac{\sigma_d^2}{K_p^2} + \frac{\pi}{2} 10^{N_i/10} B K_p, \tag{41}$$

$$\sigma_{v_g}^2 = \sigma_d^2 + \frac{\pi}{2} 10^{N_i/10} L K_p^2. \tag{42}$$

Thus with these assumptions, the cavity stability is independent of the choice of the low pass filter, which is included to limit the RF power overhead due to noise amplification.

The minimum cavity voltage variation occurs when $\frac{d\sigma_v^2}{dK_p} = 0$, which yields

$$K_p = \left(\frac{4}{\pi B} \sigma_d^2 10^{-\frac{N_i}{10}}\right)^{1/3}, \tag{43}$$

in which case,

$$\sigma_v = \frac{\sqrt{3}}{2}\left(2\pi B \sigma_d 10^{\frac{N_i}{10}}\right)^{1/3}, \text{ and} \tag{44}$$

$$\sigma_{v_g} = \sigma_d \sqrt{1 + 2\pi L \left[\frac{1}{4(\pi B \sigma_d)^2} 10^{\frac{N_i}{10}}\right]^{1/3}}. \tag{45}$$

To illustrate these results, we consider the ADC noise spectrum measured for a prototype of the LCLS-II Low Level RF (LLRF) system [2]. It is basically flat at -155 dBc/Hz above 10 kHz, and has a 1/frequency component that increases to -139 dBc/Hz at 100 Hz. However, including the 1/frequency component does not change the results significantly ($< 5\%$) for $K_p > 5000$ as the major contribution of the noise is above 10 kHz.

Fig. 10 shows the resulting rms cavity voltage stability as function of $K_p$ for $\delta = 0.615$, which corresponds to the assumed maximum peak detuning of 10 Hz in LCLS-II. Based on the measured ADC noise, $K_p$ needs to be greater than 4×10³ to meet the LCLS-II 10⁻⁴ amplitude stability (and the 0.01 degree rms phase stability). The minimum rms voltage occurs at $K_p = 3.6 \times 10^4$ as predicted by Eq. 44.

As discussed above, the choice of $L$ has phase margin implications. Fig. 11 shows $L/B$ and the associated phase margin versus $K_p$ for 1% noise related rms RF overhead. To achieve this level and a phase margin above 20 degrees, $K_p$ needs to be below 2×10⁴.

**Transient Response**

When beam is turned on after cavity has been stabilized by the feedback system, there will be a transient voltage response due to the finite bandwidth of the system. For a short period after beam turn-on ($\tau \ll T_F$), cavity detuning can be assumed constant and $e^{j\delta\tau} \approx 1$. Accordingly, the cavity voltage change in response to a beam voltage ($v_b$) step change is

$$dv \approx \frac{v_b}{2K_p}\left(2 + \frac{1-2\mu K_p - \sqrt{\Delta}}{\sqrt{\Delta}} e^{-\frac{1+\sqrt{\Delta}}{2\mu}\tau} - \frac{1-2\mu K_p + \sqrt{\Delta}}{\sqrt{\Delta}} e^{-\frac{1-\sqrt{\Delta}}{2\mu}\tau}\right), \tag{46}$$

where $\Delta = 1 - 4\mu K_p$ and $\mu = B/L$. Thus a large bandwidth is preferred to reduce the voltage overshoot.

**Conclusion**

We have derived the equations governing the cavity and generator voltages in a system where PI feedback is used to stabilize the cavity voltage in the presence of microphonics, and shown that with approximations that should be valid in most practical systems, the equations can be linearized to provide simple and accurate analytic solutions. These results should be useful for designing a feedback system when the magnitude of the detuning is reasonably well known. In particular, the maximum instrument noise can be specified to ensure that the desired cavity voltage stability is achievable.

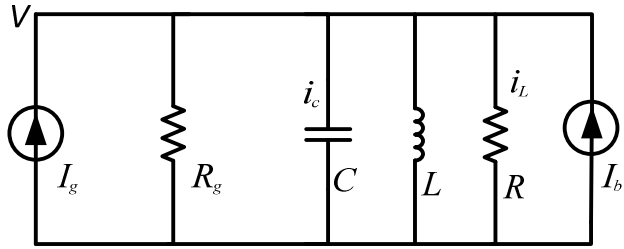

Fig. 1: An equivalent circuit model of an RF cavity.

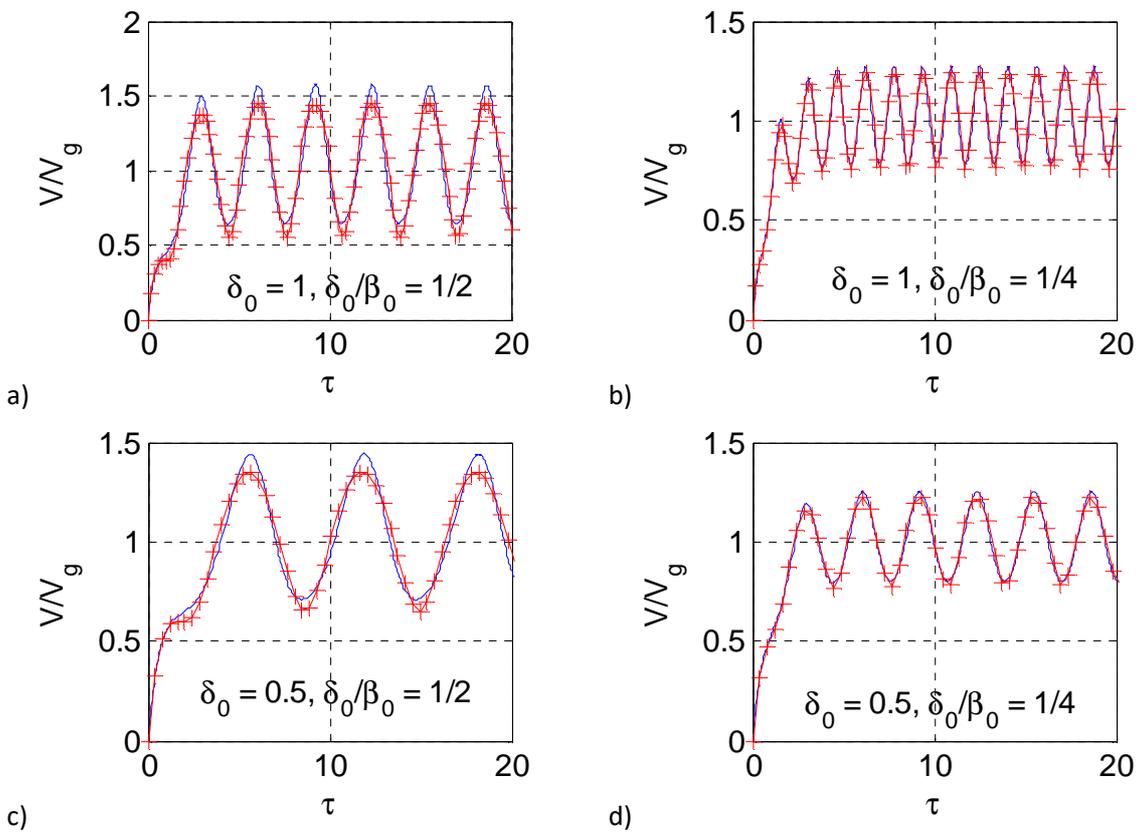

Fig. 2: Cavity voltage response to a step in $V_g$ computed numerically from the exact solution (solid lines) and to first order in $\delta_0/\beta_0$ (crosses) for various $\delta_0$ and $\delta_0/\beta_0$ values.

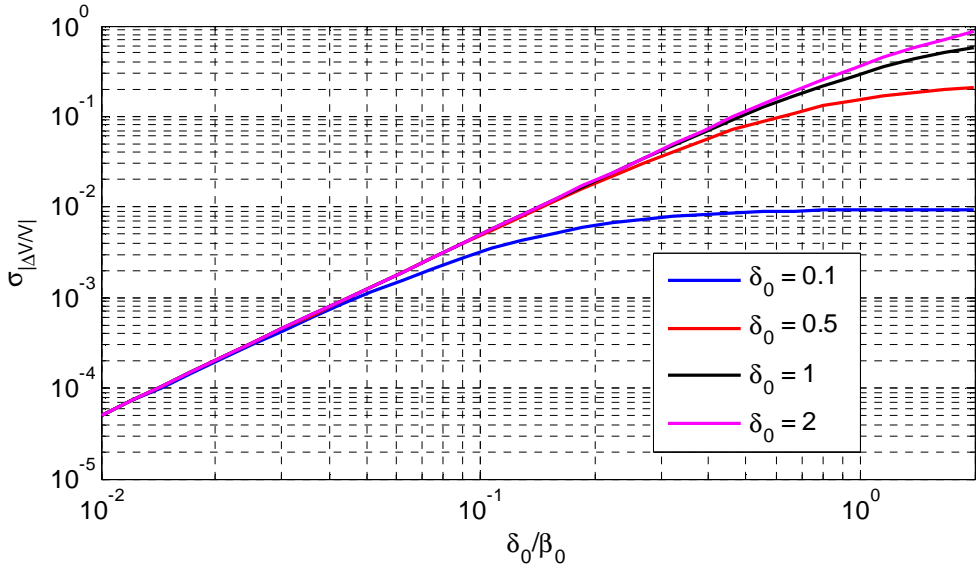

Fig. 3: RMS deviation of linear approximation from the exact numerical solution as a function of $\delta_0/\beta_0$ for different $\delta_0$.

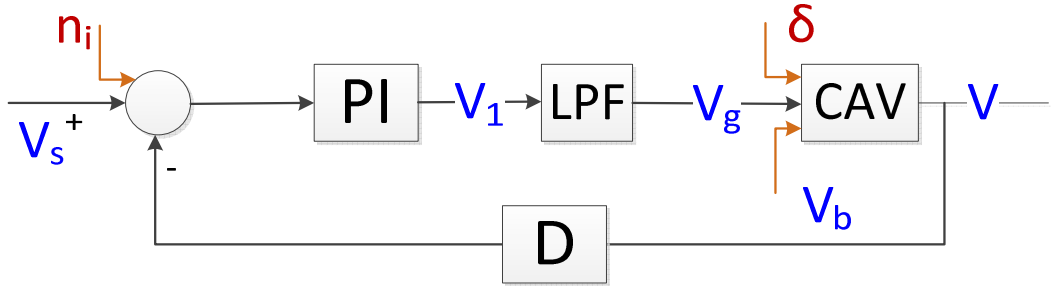

Fig. 4: Diagram of a cavity voltage controller illustrating the elements in the feedback loop: Proportional-Integral (PI) processor, low pass filter (LPF), resonant cavity (CAV) and time delay (D) through the system. $V_s$ the set point for the cavity voltage, $V_1$ is the PI processor output, $V_g$ is the generator voltage that drives the cavity, $V$ is the cavity voltage, $V_b$ is the beam loading, $n_i$ is instrument noise (e.g., from the digitizers) and $\delta$ is the cavity detuning.

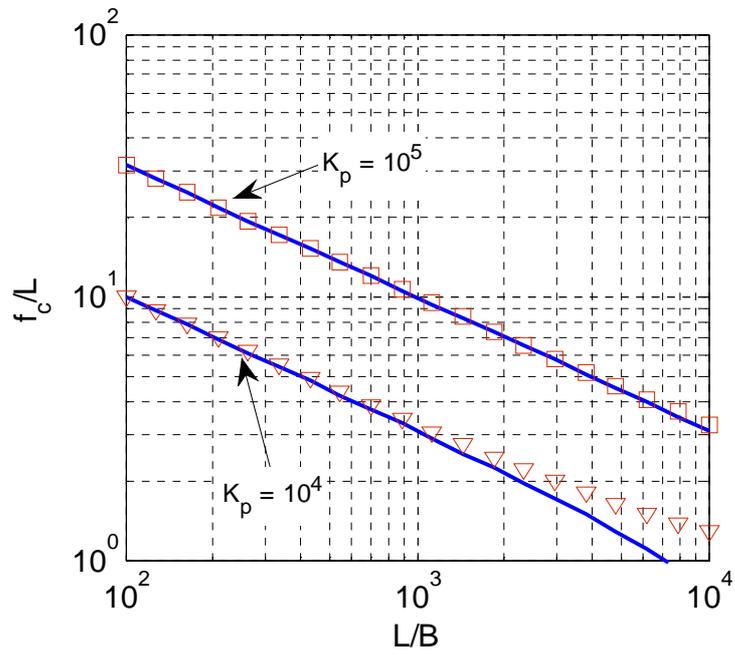

Fig. 5: Comparison of the open-loop zero-crossing frequency computed by numerical solution (solid lines) and approximation (squares and triangles).

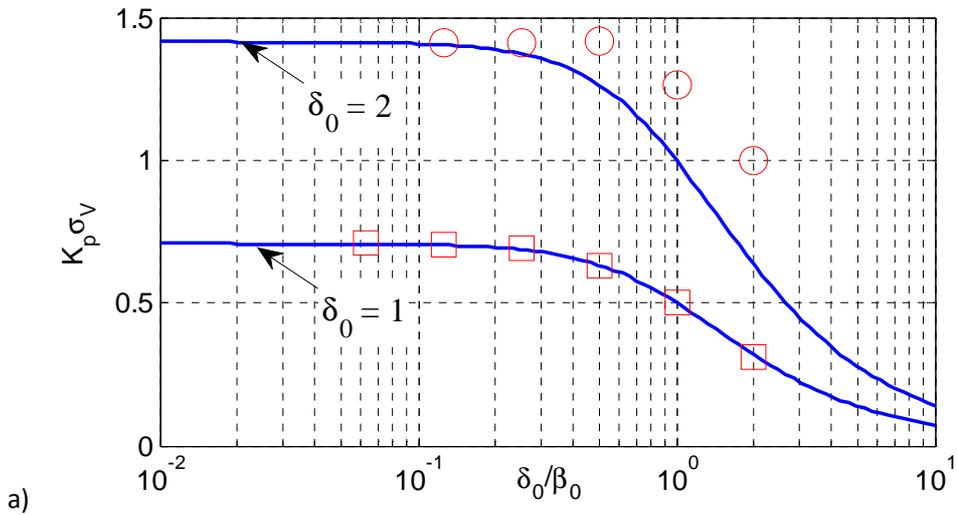

a)

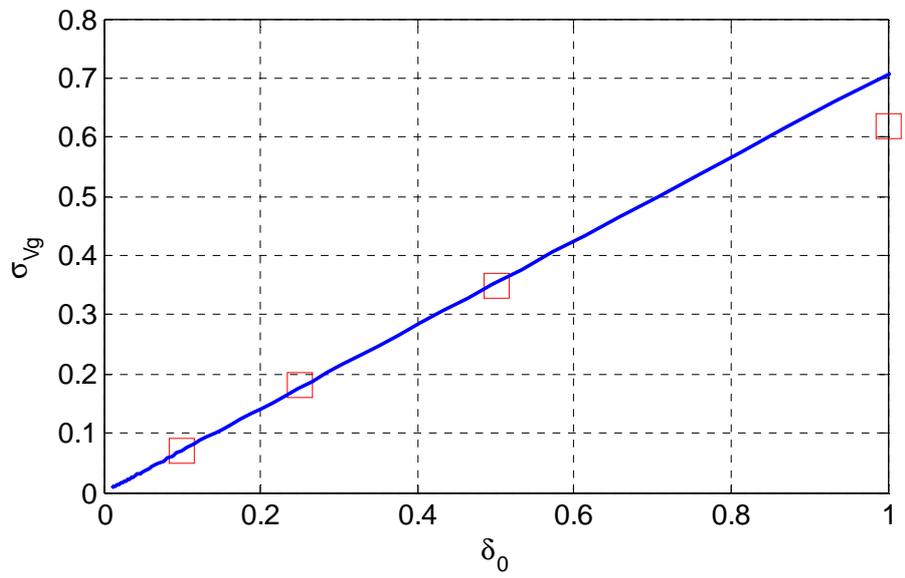

b)

Fig. 6: (a) Closed loop cavity voltage stability as a function of $\delta_0/\beta_0$ for two values of $\delta_0$ and (b) the generator variation as a function of $\delta_0$. The solid line and squares are, respectively, the numerical solution to Eq. 16 and the linear approximation (Eq. 32 and Eq. 33). For the former, $K_p$ = 1000 was assumed for (a) and $\beta_0$ = 600 for (b).

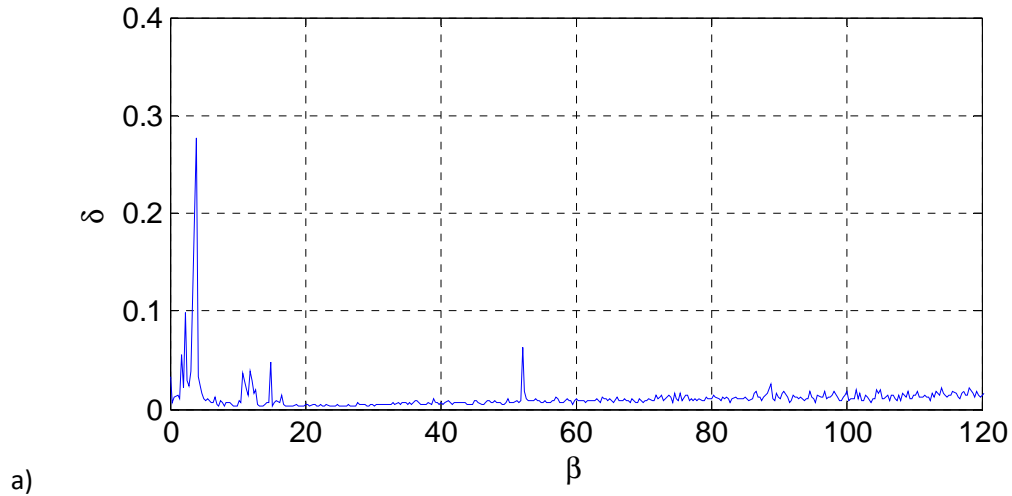

a)

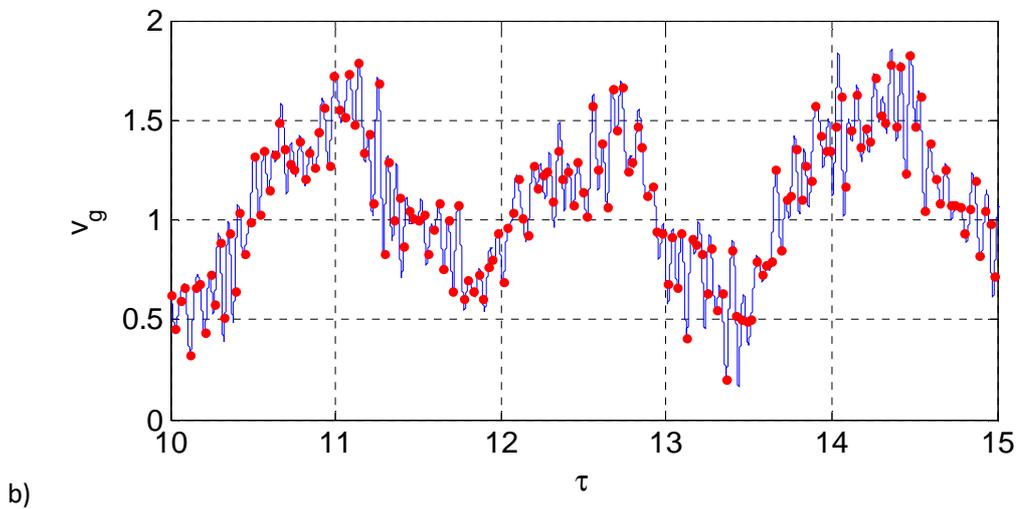

b)

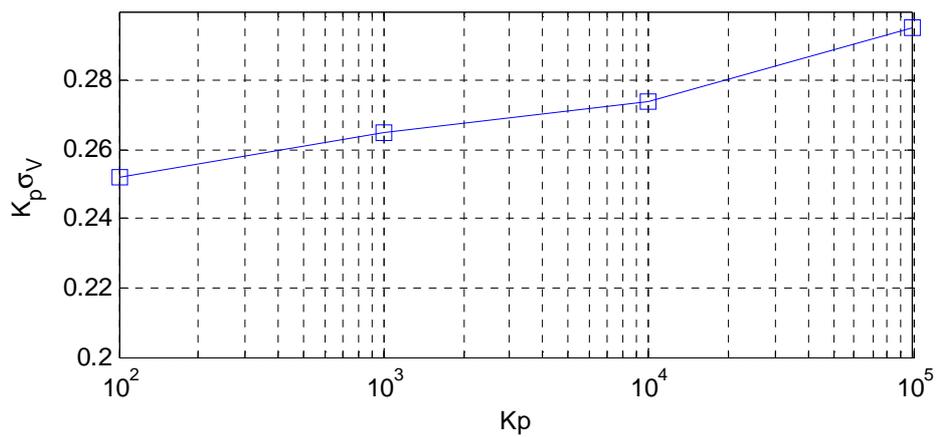

c)
Fig. 7: (a) Detuning spectrum measured at FNAL for an LCLS-II cavity, (b) the corresponding simulated closed loop generator voltage where the solid line and dots are, respectively, the numeric solution to Eq. 16 and linear approximation computed from Eq. 21 and Eq. 31, and c) $K_p$ times the cavity voltage stability, computed by solving Eq. 16, versus $K_p$.

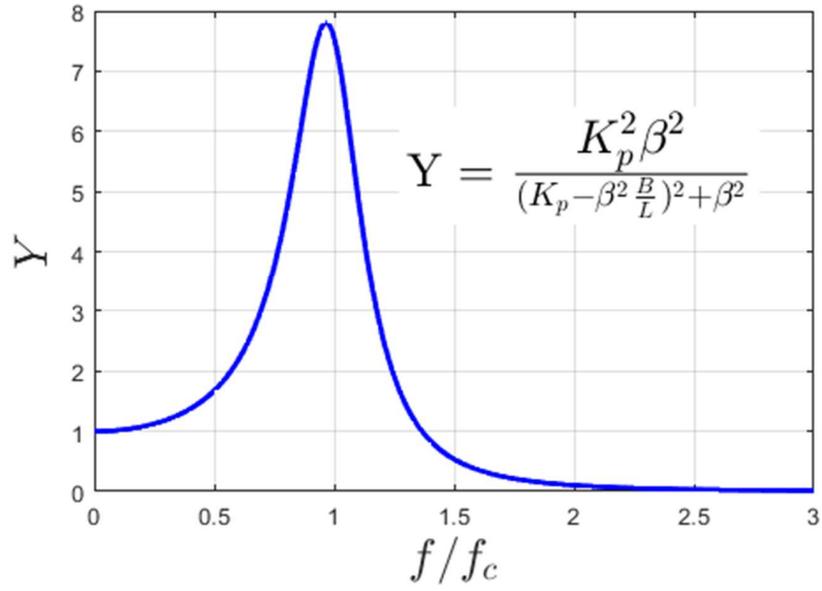

a)

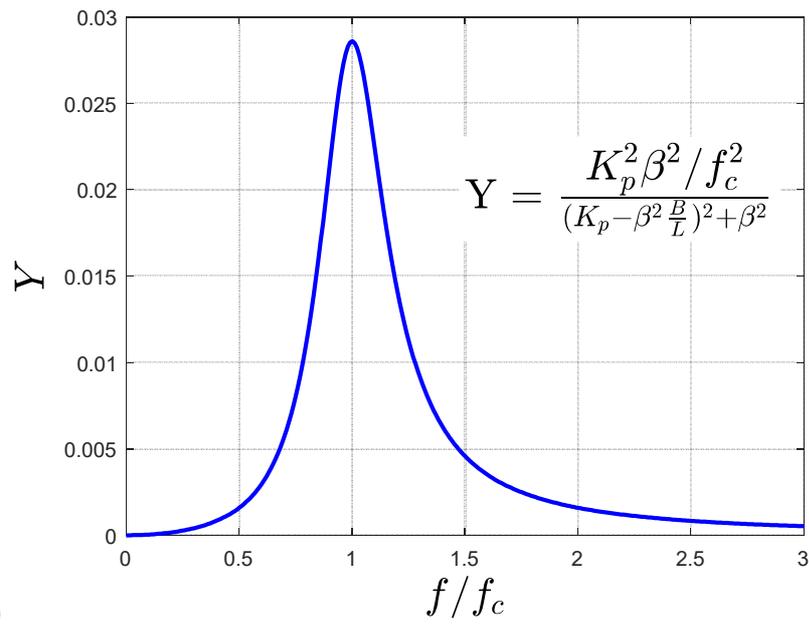

b)

Fig. 8: (a) Cavity and (b) drive weighting functions in Eq. 37 and Eq. 38, respectively, for a phase margin of 20 degrees. These curves are independent of $K_p$.

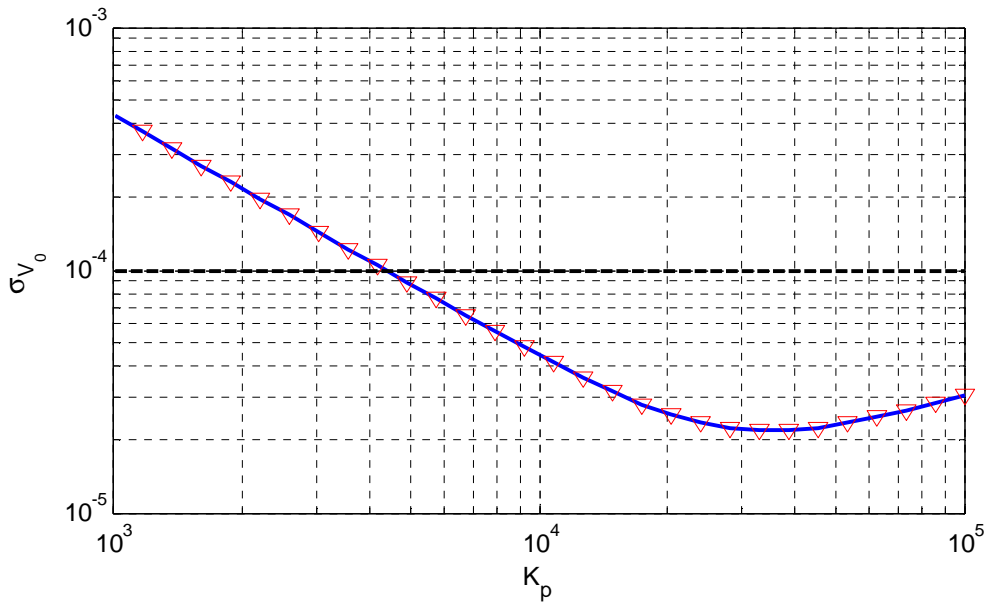

Fig. 9: The rms cavity voltage stability versus $K_p$ for δ = 0.615 where the solid line and triangles assume, respectively, the measured ADC noise spectrum and a flat -155 dBc/Hz spectrum.

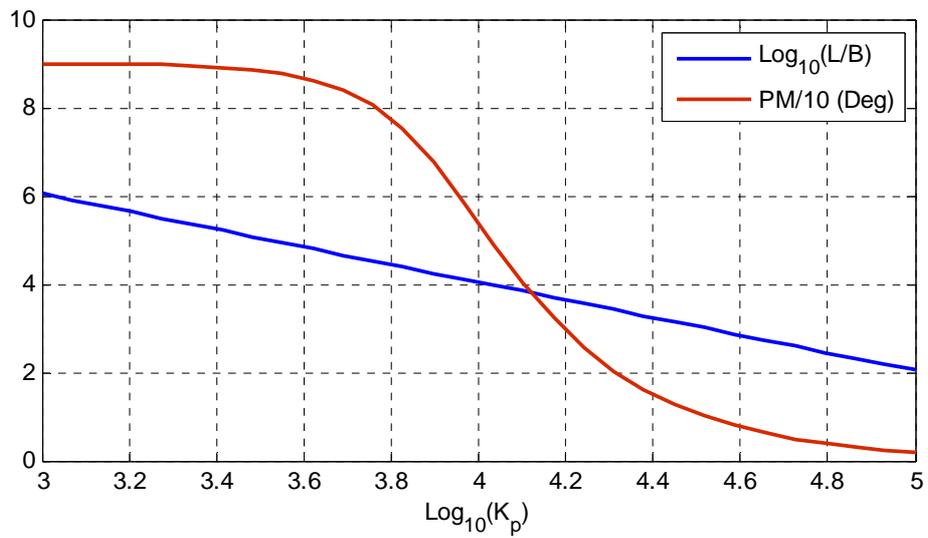

Fig.10: $L/B$ and the corresponding phase margin as a function of $K_p$ for 1% noise related RF overhead.